\documentclass{article}
\usepackage{spconf,amsmath,graphicx,hyperref}

\usepackage{times}
\usepackage{epsfig}
\usepackage{amssymb}
\usepackage{booktabs}
\usepackage[acronym]{glossaries}
\usepackage{fancyhdr}
\usepackage{multicol}
\usepackage{nicefrac}
\usepackage{transparent}
\usepackage{watermark}
\usepackage{pifont}

\usepackage{booktabs}
\usepackage{multirow}
\usepackage{array}
\usepackage{caption}
\usepackage{siunitx}
\usepackage{algorithm, algorithmic}
\usepackage{makecell}
\usepackage{tabularx}   % add in preamble
\usepackage{siunitx}

\usepackage{xcolor}
\definecolor{lightblue}{RGB}{173,216,230} % pastel light blue

\usepackage{fancyhdr}
\fancypagestyle{firstpage}{
  \fancyhf{}
  \fancyfoot[C]{%
    \parbox{\textwidth}{\centering\footnotesize
    \copyright\ 2026 IEEE. Personal use of this material is permitted.
    Permission from IEEE must be obtained for all other uses, in any current or future media,
    including reprinting/republishing this material for advertising or promotional purposes,
    creating new collective works, for resale or redistribution to servers or lists,
    or reuse of any copyrighted component of this work in other works.}
  }

}

\newcolumntype{P}[1]{>{\raggedleft\arraybackslash}p{#1}}

\usepackage[draft]{pdfcomment} % https://ctan.org/pkg/pdfcomment?lang=en 

\usepackage{xspace}

\glsdisablehyper % to disable hyperlinks on all acronyms

\definecolor{Sepia}{RGB}{112,66,20}
\definecolor{AcColor}{rgb}{0,0,0}

\newcommand{\accolor}[1]{\textcolor{AcColor}{#1}} % subsequent uses
\newacronymstyle{myacro}
{%
  \GlsUseAcrEntryDispStyle{long-short}%
}%
{%
  \GlsUseAcrStyleDefs{long-short}%
}

\setacronymstyle{myacro}

\let\xx\tip

\newacronym{adc}{ADC}{Analog to Digital Converter}
\newacronym{admm}{ADMM}{Alternating Direction Method of Multipliers}
\newacronym{asic}{ASIC}{Application Specific Integrated Circuit}
\newacronym{cem}{CEM}{Cross-Entropy Method}
\newacronym{cots}{COTS}{Commodity Off-The-Shelf}
\newacronym{cpu}{CPU}{Central Processing Unit}
\newacronym{cnn}{CNN}{Convolutional Neural Network}
\newacronym{ddp}{DDP}{Differential Dynamic Programming}
\newacronym{dnn}{DNN}{Deep Neural Network}
\newacronym{dns}{DNS}{Deep Noise Suppression}
\newacronym{dof}{DOF}{Degree Of Freedom}
\newacronym{dram}{DRAM}{Dynamic RAM}
\newacronym{fc}{FC}{Fully Connected}
\newacronym{fpga}{FPGA}{Field Programmable Gate Array}
\newacronym{gpu}{GPU}{Graphics Processing Unit}
\newacronym{gru}{GRU}{Gated Recurrent Unit}
\newacronym{hil}{HIL}{Hardware In the Loop}
\newacronym{ip}{IP}{Interior Point}
\newacronym{ipb}{IP block}{Intellectual Property Block}
\newacronym{lstm}{LSTM}{Long Short Term Memory}
\newacronym{ltv}{LTV}{Linear Time Varying}
\newacronym{mac}{MAC}{Multiply-Accumulate}
\newacronym{mae}{MAE}{Median Angular Error}
\newacronym{mlp}{MLP}{Multilayer Perceptron}
\newacronym{mee}{MEE}{Median Endpoint Error}
\newacronym{mpc}{MPC}{Model Predictive Control}
\newacronym{mpcer}{MPC}{Model Predictive Controller}
\newacronym{mppi}{MPPI}{Model Predictive Path Integral}
\newacronym{nc}{NC}{Neural Controller}
\newacronym{nmpc}{NMPC}{Nonlinear Model Predictive Control}
\newacronym{nn}{NN}{Neural Network}
\newacronym{npc}{NPC}{Neural Predictive Control}
\newacronym{npu}{NPU}{Neural Processing Unit}
\newacronym{ode}{ODE}{Ordinary Differential Equation}
\newacronym{pcb}{PCB}{Printed Circuit Board}
\newacronym{pd}{PD}{Proportional Derivative}
\newacronym{pid}{PID}{Proportional Integral Derivative}
\newacronym{pl}{PL}{Programmable Logic}
\newacronym{ps}{PS}{Processing System}
\newacronym{pso}{PSO}{Particle Swarm Optimization}
\newacronym{pwm}{PWM}{Pulse Width Modulation}
\newacronym{qp}{QP}{Quadratic Programming}
\newacronym{rl}{RL}{Reinforcement Learning}
\newacronym{rnn}{RNN}{Recurrent Neural Network}
\newacronym{ros}{ROS}{Robot Operating System}
\newacronym{rpgd}{RPGD}{Resampling Parallel Gradient Descent}
\newacronym{slp}{SLP}{Single Layer Perceptron}
\newacronym{sm}{SM}{Supplementary Material}
\newacronym{snr}{SNR}{Signal-to-Noise Ratio}

\newacronym[description={System on Chip; FPGA with embedded programmable processor}]{soc}{SoC}{System on Chip}
\newacronym{sqp}{SQP}{Sequential Quadratic Programming}
\newacronym{sram}{SRAM}{Static RAM}
\newacronym{usb}{USB}{Universal Serial Bus}
\newacronym{vga}{VGA}{Video Graphics Adaptor}
\newacronym{uart}{UART}{Universal Asynchronous Receiver/Transmitter}
\newacronym{xla}{XLA}{Accelerated Linear Algebra}

\newacronym{ssm}{SSM}{State Space Model}
\newacronym{film}{FiLM}{Feature-wise Linear Modulation}
\newacronym{dprnn}{DPRNN}{Dual-Path RNN}
\newacronym{sota}{SOTA}{State Of The Art}
\newacronym{ec}{EC}{Embedding Concatenation}
\newacronym{ssmm}{SSMM}{SSM Modulation}
\newacronym{sisnr}{SISNR}{Scale-Invariant Signal-to-Noise Ratio}
\newacronym{lr}{lr}{learning rate}
\newacronym{se}{SE}{Speech Enhancement}
\newacronym{ola}{OLA}{Overlap-and-Add}
\newacronym{anc}{ANC}{Active Noise Control}
\newacronym{sisdr}{SI-SDR}{Scale Invariant Signal-to-Distortion Ratio}
\newacronym{mse}{MSE}{Mean Squared Error}
\newacronym{ood}{OOD}{Out-Of-Distribution}
\newacronym{lora}{LoRA}{Low-rank Adaptation}
\newacronym{erb}{ERB}{Equivalent Rectangular Bandwidth}
\newacronym{stft}{STFT}{Short-Time Fourier Transform}

\title{Towards Lightweight Adaptation of \\Speech Enhancement Models in Real-World Environments }
\name{Longbiao Cheng, Shih-Chii Liu\thanks{This work was partially funded by the Swiss National Science Foundation projects CA-DNNEdge (208227).}}
\address{Institute of Neuroinformatics, University of Zurich and ETH Zurich}
\begin{document}
\ninept
\maketitle
\thispagestyle{firstpage}

\begin{abstract}
Recent studies have shown that post-deployment adaptation can improve the robustness of speech enhancement models in unseen noise conditions. However, existing methods often incur prohibitive computational and memory costs, limiting their suitability for on-device deployment. 
In this work, we investigate model adaptation in realistic settings with dynamic acoustic scene changes and propose a lightweight framework that augments a frozen backbone with low-rank adapters updated via self-supervised training. 
Experiments on sequential scene evaluations spanning 111 environments across 37 noise types and three signal-to-noise ratio ranges, including the challenging [-8,0]~dB range, show that our method updates fewer than 1\% of the base model's parameters while achieving an average 1.51~dB SI-SDR improvement within only 20 updates per scene. 
Compared to state-of-the-art approaches, our framework achieves competitive or superior perceptual quality with smoother and more stable convergence, demonstrating its practicality for lightweight on-device adaptation of speech enhancement models under real-world acoustic conditions.
\end{abstract}


\section{Introduction}
\label{sec:intro}

\xx{se} %is a fundamental function in hearing instruments and %related assistive devices such as hearing aids, cochlear implants, and augmented reality glasses, where it 
plays an important role in improving intelligibility and reducing listening effort for listeners using hearing instruments and other hearing assistive devices  %for both hearing-impaired and normal-hearing listeners 
in noisy environments~\cite{wendt2017impact, wang2017deep, cheng2025modulating, cheng2024dynamic}. Recent neural network–based approaches% have achieved remarkable progress, with substantial 
have demonstrated substantial gains across a wide range of acoustic conditions~\cite{wang2018supervised,o2024speech}. Despite these advances, a challenge remains: the limited generalizability of deep learning models. Systems that perform well under training conditions often suffer degradation when deployed in mismatched environments, such as in the presence of unseen noise types, mismatched microphones, or different speech characteristics.

To %address this challenge, 
improve the generalizability of networks, 
extensive data augmentation is often employed during training, for example, by incorporating diverse noise sources and speaker populations~\cite{braun2020data, chen2024improving}. In parallel, more advanced model architectures have been developed, including self-supervised learning backbones pretrained on large-scale audio datasets~\cite{huang2022investigating, hung2022boosting, lee2024leveraging} and generative models such as diffusion-based approaches that explicitly model speech and noise distributions~\cite{gonzalez2024investigating, richter2024causal, richter2024diffusion}. Although  these methods can improve generalizability of networks, they typically come at the cost of  increased model complexity, therefore making it difficult to deploy these models for the edge.  %deployment on resource-constrained devices difficult.

% (starts with solving the ood data problem by continual learning). similar concept has also been explored on adapting se models after deployment. 
Since lightweight models are inherently limited in their capacity to handle a wide range of acoustic scenarios, an alternative method would be exploring continual learning~\cite{hu2019overcoming,ott2023biologically} for post-deployment adaptation. 
%In this setting, a pretrained model is adapted to the target environment using only noisy recordings collected in situ. 
RemixIT-based approaches~\cite{tzinis2022remixit, fang2024uncertainty,han2024unsupervised,liao2025leveraging}, for example, leverage the pretrained network to generate pseudo targets for training a student network better suited to the deployed environment. More recent work has explored test-time training, where a noisy-target auxiliary task is introduced to update shared backbone layers during adaptation~\cite{behera2025test}. However, these methods often require a significant increase in parameters to support teacher–student frameworks or auxiliary tasks and involve fine-tuning a large percentage  of the pretrained model parameters. 
Besides  imposing memory and computational requirements beyond that available on low-resource platforms, convergence during network training also takes longer.

% Furthermore, prior adaptation studies focus on improving the enhancement performance on \xx{ood} datasets. Such datasets usually contain a broad mixture of speaker and noise conditions and wide \xx{snr} ranges. In contrast, real-world deployment often involves harsher and more constrained conditions, such as consistently low \xx{snr} ranges and limited speakers and persistent noise, where adaptation is considerably more challenging yet critical for usability.  

Furthermore, most prior adaptation studies aim at improving \xx{se} performance on \xx{ood} datasets, which typically include a broad mixture of speakers, noise types, and wide \xx{snr} ranges. While such datasets are useful for benchmarking, they differ significantly from real world environments. In practice, acoustic conditions are organized into scenes, and at any given time the system operates within a single scene. Over time, however, scenes evolve, leading to sequential scene changes. Addressing the adaptation process under such scene changes, rather than adapting once to a diverse but static dataset, presents a more demanding yet practically essential challenge for achieving robust usability.

% In this work, we address the problem of lightweight and stable \xx{se} model adaptation under realistic acoustic conditions. Our main contributions are:  
% \begin{itemize}
%     \item We formalize the adaptation setting for \xx{se} models under real-world acoustic constraints that involving scene changes.
%     \item We propose a lightweight self-supervised adaptation framework that avoids full model fine-tuning by introducing low-rank adapters~\cite{hu2022lora}.  
%     % \item We introduce a metric for quantifying update stability during adaptation. 
%     \item Through experiments with multiple \xx{se} backbones across 111 noisy environments, we show that updating less than 1\% of the pretrained parameters already yields a \xx{sisdr} improvement of 1.13 dB with only 20 adaptation steps, even under highly noisy conditions where the \xx{snr} ranges from -8 to 0~dB.  
% \end{itemize} 
% These findings highlight the potential for practical on-device lightweight adaptation of edge \xx{se} models in real-world scenarios.  
In this work, we address the problem of lightweight \xx{se} model adaptation under realistic acoustic conditions. The  contributions of our work are as follows:  
\begin{itemize}
    \item We formalize an adaptation setting for \xx{se} models in real-world acoustic scenarios, where the acoustic scene changes over time.  
    \item We propose a lightweight self-supervised adaptation framework based on low-rank adapters~\cite{hu2022lora}, avoiding the need to fine-tune the full model.  
    \item We evaluate the proposed framework on two \xx{se} backbones across 111 noisy environments, showing an average 1.51~dB improvement in \xx{sisdr} within only 20 adaptation steps for a scene, while updating fewer than 1\% of the pretrained parameters.  
\end{itemize}  
These results demonstrate the practicality of lightweight adaptation for robust \xx{se} in dynamic, real-world acoustic environments.

\section{Problem Formulation}
\label{sec:problem}

We consider the single-channel \xx{se} task, where the observed signal $y(t)$ is modeled as
\begin{equation}
y(t) = s(t) + n(t),
\end{equation}
with $s(t)$ denoting the clean speech signal and $n(t)$ denoting additive interference. 
The goal of \xx{se} is to estimate $\hat{s}(t)$ that closely approximates $s(t)$ from the noisy input $y(t)$.

Let $f_{\theta}$ denote a deep neural network trained on a \emph{source} corpus of paired noisy and clean speech, with $\theta_{0}$ representing the parameters of the pretrained base model. At deployment, the system will operate in different acoustic \emph{scenes}, where each scene corresponds to a relatively stable combination of factors such as background noise type, speakers, and \xx{snr} range. We denote the adaptation dataset from a given scene as
\begin{equation}
\mathcal{D}_{\text{adapt}}^{(m)} = \{ y_i^{(m)} \}_{i=1}^{N_m},
\end{equation}
where $m$ indexes the scene and no clean references are available. The objective is to update the model parameters to $\theta_{m}$ such that the adapted model $f_{\theta_m}$ achieves improved enhancement performance in scene $m$, as evaluated on a disjoint test set $\mathcal{D}_{\text{test}}^{(m)}$ collected under the same conditions.

In realistic usage, however, the acoustic environment is not fixed. Over time, the system will encounter scene changes, where the acoustic conditions shift from $m$ to a new scene $m+1$ (e.g., a new noise source, a different \xx{snr}, or change in speakers). The model must therefore adapt continually, updating from $\theta_{m}$ to $\theta_{m+1}$ using only the data from $\mathcal{D}_{\text{adapt}}^{(m+1)}$. This continual adaptation under scene changes introduces additional challenges beyond one-time adaptation to an \xx{ood} dataset, where all scenes are available simultaneously during adaptation and the model can optimize jointly across them.

\section{Methods}
\label{sec:adpt}

Since noisy–clean audio pairs are not available in the adaptation setting formulated in Sec.~\ref{sec:problem}, conventional supervised fine tuning cannot be applied. We therefore propose a self supervised framework that leverages the pretrained base model $f_{\theta_{0}}$ to generate pseudo targets, and employs lightweight low rank adapters to enable efficient and stable updates as scenes evolve.

\subsection{Self-Supervised Adaptation Framework}

Following previous works~\cite{tzinis2022remixit, fang2024uncertainty,han2024unsupervised,liao2025leveraging}, the adaptation procedure begins by using the pretrained base model $f_{\theta_{0}}$ to generate a pseudo clean estimate from a noisy input $y \in \mathcal{D}_{\text{adapt}}^{(m)}$ of scene $m$:  
\begin{equation}
\hat{x} = f_{\theta_{0}}(y).
\end{equation}
This estimate $\hat{x}$ serves as the teacher target during adaptation. To construct a self supervised learning signal, we sample a separate noise segment $n \in \mathcal{D}_{\text{adapt}}^{(m)}$ from the same scene, scale it by a factor $\alpha$ corresponding to a randomly sampled \xx{snr}, and re-mix it with $\hat{x}$ to obtain the adaptation input:
\begin{equation}
\tilde{y} = \hat{x} + \alpha n.
\end{equation}
The adapted model $f_{\theta_{m}}$ then produces an enhanced output:
\begin{equation}
\tilde{x} = f_{\theta_{m}}(\tilde{y}).
\end{equation}
The parameters $\theta_{m}$ are updated by minimizing a task specific loss $\mathcal{L}(\tilde{x}, \hat{x})$ between the adapted output and the pseudo target:
\begin{equation}
\theta_{m} = \arg\min_{\theta} \, \mathbb{E}_{y \sim \mathcal{D}_{\text{adapt}}^{(m)}} \Big[ \mathcal{L}\big( f_{\theta}(\tilde{y}), \hat{x} \big) \Big].
\end{equation}

When a scene change occurs, the same procedure is repeated with the new dataset $\mathcal{D}_{\text{adapt}}^{(m+1)}$, and the model parameters will be adapted from $\theta_{m}$ to $\theta_{m+1}$. In this way, the system  will be able to track a sequence of evolving conditions by adapting step by step across scenes. 

\subsection{Low-Rank Adapters}

A straightforward adaptation strategy is to fine tune all parameters $\theta$. While this may yield improvements within a single scene, it is prone to overfitting and often suffers from catastrophic forgetting of the knowledge encoded in the base model $f_{\theta_{0}}$, which normally is trained on a comprehensive source dataset covering diverse speakers, noise types, and \xx{snr} conditions.

To overcome these challenges, we adopt \xx{lora}~\cite{hu2022lora}, which restricts adaptation to a low dimensional subspace while keeping the pretrained backbone frozen. For a pretrained weight matrix $W_{0} \in \mathbb{R}^{d \times k}$ in $\theta_{0}$, the scene specific parameterization is defined as  
\begin{equation}
\label{eq:lora}
W_{m} = W_{0} + \beta B_m A_m, 
\qquad 
B_m \in \mathbb{R}^{d \times r}, \quad A_m \in \mathbb{R}^{r \times k},
\end{equation}
where $m$ denotes the current scene, $r \ll \min(d,k)$ is the adaptation rank, and $\beta$ is a scaling factor.  
During adaptation within scene $m$, only $A_m$ and $B_m$ are updated, while the base weights $W_{0}$ remain fixed. At inference time, the adapted model is obtained by merging the residual update $B_m A_m$ with the base weights. The adaptation process is summarized in Alg.~\ref{alg:adapt}.

This adaptation method is naturally suited to continuous adaptation across scene changes. The performance of the \xx{se} model under a specific scene $m$ is enhanced by its own lightweight adapter pair $(A_m, B_m)$, while preserving the general knowledge encoded in $W_{0}$. When the acoustic conditions shift from scene $m$ to scene $m+1$, the system only needs to transition from $(A_{m}, B_{m})$ to a new pair $(A_{m+1}, B_{m+1})$, without modifying the backbone. 
In this way, adaptation remains localized to scene specific parameters, ensuring efficiency and mitigating catastrophic forgetting. 

\begin{algorithm}[t]
\caption{Self-Supervised Low-Rank Adaptation }
\label{alg:adapt}
\begin{algorithmic}[1]
\REQUIRE $f_{\theta_{0}}$: Frozen base model;  $\phi_m = \{A_m,B_m\}$: Adapter parameters for scene $m$
\FOR{each noisy utterance $y$ in the current scene $m$}
    \STATE Generate pseudo target: $\hat{x} = f_{\theta_{0}}(y)$
    \STATE Sample a noise segment $n$ and scaling $\alpha$ according to target \xx{snr}
    \STATE Create adaptation input: $\tilde{y} = \hat{x} + \alpha n$
    \STATE Forward adapted model: $\tilde{x} = f_{\theta_{0},\phi_m}(\tilde{y})$
    \STATE Update $\phi_m$ by minimizing $\mathcal{L}(\tilde{x}, \hat{x})$
\ENDFOR
\STATE For evaluation in scene $m$, merge $\phi_m$ into the base model via Eq.~\ref{eq:lora}
\end{algorithmic}
\end{algorithm}

\begin{table*}[t]
% \vspace{-6pt}
\setlength{\tabcolsep}{4.25pt}
\centering
\caption{Comparison of the proposed adaptation framework and RemixIT framework for GRU and DPRNN models in isolated (\ding{55}) and sequential (\ding{51}) scene settings. 
\textbf{Bold} values mark the best performance within each model configuration, while \textcolor{blue}{blue} values indicate  performance degradation relative to the pretrained baseline.}
\vspace{-2mm}
\begin{tabular}{l c l r l ccc ccc ccc}
\toprule
\multirow{2}{*}[-0.5ex]{Model} 
& \multirow{2}{*}[-0.5ex]{\makecell{Sequential \\ Scene?}} 
& \multirow{2}{*}[-0.5ex]{\makecell{Adaptation \\ Framework}} 
& \multicolumn{2}{c}{Adaptable Params} 
& \multicolumn{3}{c}{SNR $\in$ [-8, 0] dB} 
& \multicolumn{3}{c}{SNR $\in$ [0, 5] dB} 
& \multicolumn{3}{c}{SNR $\in$ [5, 10] dB} \\ 
\cmidrule(lr){4-5} \cmidrule(lr){6-8} \cmidrule(lr){9-11} \cmidrule(lr){12-14}
& &
& \multicolumn{1}{r}{\#} & \% 
& PESQ & STOI & SI-SDR 
& PESQ & STOI & SI-SDR 
& PESQ & STOI & SI-SDR \\
\midrule \midrule

\multicolumn{5}{c}{Noisy Speech}
& 1.08 & 67.88 & -3.89 
& 1.16 & 81.48 & 2.53 
& 1.35 & 89.25 & 7.49 \\\midrule

\multirow{5}{*}[-1.5ex]{GRU*} & \multicolumn{4}{c}{Pretrained} 
& 1.16 & 71.01 & 3.86 
& 1.35 & 84.85 & 6.82 
& 1.57 & 90.99 & 9.58 \\ \cmidrule(lr){2-14} 

& \ding{55} & RemixIT & 230,144 & 100 
& 1.19 & 71.64 & 4.64 
& 1.39 & 84.96 & 8.27 
& 1.62 & 91.07 & 11.50 \\

& \ding{55} & Ours & \textbf{512} & \textbf{0.22} 
& 1.22 & 71.96 & 4.71 
& 1.44 & 85.34 & 8.36 
& 1.67 & 91.38 & 11.43 \\ \cmidrule(lr){2-14} 

& \ding{51} & RemixIT & 230,144 & 100 
& 1.18 & \textcolor{blue}{70.34} & 4.63 
& \textcolor{blue}{1.34} & \textcolor{blue}{83.13} & 8.42 
& \textcolor{blue}{1.51} & \textcolor{blue}{88.63} & 11.03 \\

& \ding{51} & Ours &  \textbf{512} & \textbf{0.22}
& \textbf{1.23}	& \textbf{72.65} & \textbf{4.84}	
& \textbf{1.47} & \textbf{85.84} & \textbf{8.65}	
& \textbf{1.72}	& \textbf{91.64} & \textbf{11.89}	 \\ \midrule

\multirow{5}{*}[-1.5ex]{DPRNN*} & \multicolumn{4}{c}{Pretrained} 
& 1.21 & 74.93 & 5.15 
& 1.46 & 87.82 & 8.70 
& 1.73 & 93.05 & 11.91 \\  \cmidrule(lr){2-14} 

& \ding{55} & RemixIT & 89,258 & 100 
& 1.24 & \textcolor{blue}{73.74} & 5.58 
& 1.49 & \textcolor{blue}{87.48} & 9.88 
& 1.79 & \textcolor{blue}{92.91} & 13.45 \\ 

& \ding{55} & Ours & \textbf{708} & \textbf{0.79} 
& 1.26 & \textbf{75.52} & 5.52 
& 1.54 & \textbf{87.93} & 9.46 
& 1.84 & 93.19 & 12.94 \\  \cmidrule(lr){2-14} 

& \ding{51} & RemixIT & 89,258 & 100 
& \textbf{1.27} & \textcolor{blue}{74.51} & 5.82
& \textcolor{blue}{1.44} & \textcolor{blue}{87.68} & \textbf{10.11}	
& \textcolor{blue}{1.66} & \textcolor{blue}{92.78} & 13.60 \\

& \ding{51} & Ours & \textbf{708} & \textbf{0.79} 
&\textbf{1.27}	& 75.18	& \textbf{5.85}	
& \textbf{1.54}	& \textbf{87.93}	& \textbf{10.11}	
& \textbf{1.84}	& \textbf{93.21}	& \textbf{13.76} \\

\bottomrule
\multicolumn{10}{l}{*: GRU model: 230k parameters, 17M MAC/s; DPRNN model: 89k parameters, 1503M MAC/s.} \\

\end{tabular}

\label{tab:adapt_results}
\end{table*}

\begin{figure*}[t]
  \centering
  \includegraphics[width=2.075\columnwidth]{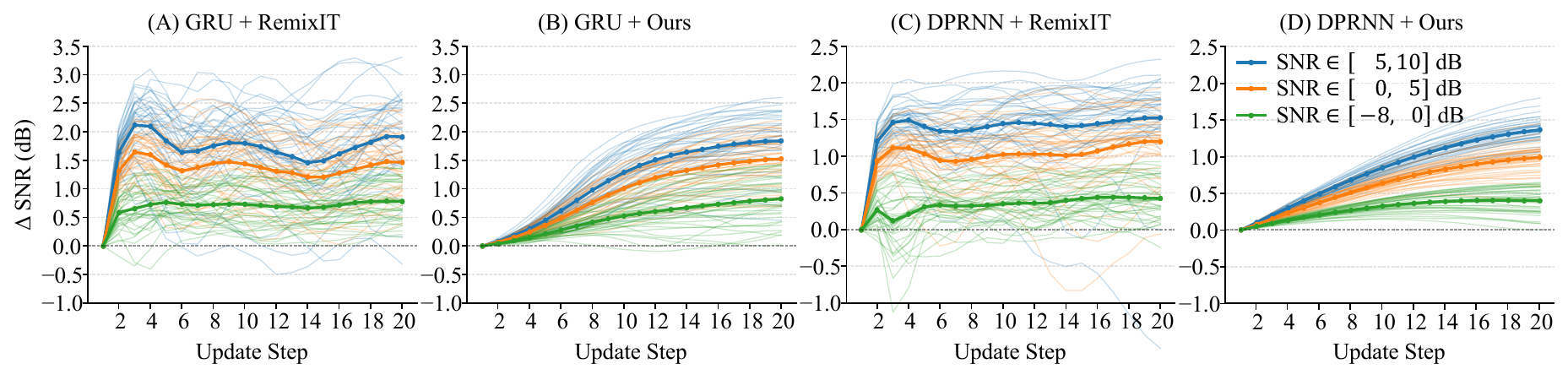}  % adjust path if needed
  \vspace{-5mm}
  \caption{Per-update SNR improvement ($\Delta$SNR in dB) of adapted GRU and DPRNN backbones across three SNR ranges. Light curves indicate individual acoustic scenes. Adaptation with RemixIT achieves rapid early gains but exhibits unstable trajectories, while our method provides steady and consistent improvements over steps.}
  \vspace{-4mm}

\label{fig:snr_improvement}
\end{figure*}

\section{Experiment Setups}

\subsection{Networks}

We evaluate the proposed methods using two representative \xx{se} architectures: a \xx{gru}-based network and a DPRNN-based network~\cite{luo2020dual}.

The \xx{gru} network consists of an input \xx{fc} layer with 128 units, followed by two \xx{gru} layers with 128 hidden states each, and an output FC layer with 128 units. The input features are compressed \xx{erb}-domain magnitude spectrograms, extracted with 128 \xx{erb} filters and a compression ratio of 0.3. The analysis uses a frame length of 512 and a hop size of 256. The network predicts a 128-dimensional \xx{erb} mask, with values constrained to [0,1] by a sigmoid activation. The resulting network has 230.14 k parameters and a computation cost of 16.80 M \xx{mac} per second. During adaptation, \xx{lora} are performed on both the input and output \xx{fc} layers, with a rank of 1 and a scaling factor of 64. 

The DPRNN network operates directly in the time-frequency domain, where the input is the concatenation of the real and imaginary parts of the \xx{stft} spectrogram, computed with a frame length of 320 and a hop size of 160. The model estimates a complex mask, with values constrained to [-1, 1] using a tanh activation. The architecture begins with an input convolutional layer that maps the 2-dimensional input (real and imaginary) to 32 channels. The network contains four dual-path blocks. Each block includes: 1) an inter-frame module, implemented with a single unidirectional \xx{gru} with 32 units followed by an \xx{fc} layer with 32 units, and 2) an intra-frame module, implemented with a bidirectional \xx{gru} with 32 units per direction, followed by an \xx{fc} layer with 32 units.
The output is produced by a convolutional layer that maps the 32-channel features back to 2 channels (real and imaginary). The DPRNN model has a parameter size of 89.25k at a computation cost of 1503.32 M \xx{mac} per second.
Since both input and output convolutional layers in the DPRNN use a kernel size of 1, they are implemented using \xx{fc} layers. During adaptation, \xx{lora} are applied to all \xx{fc} layers, with a rank of 1 and  scaling factor of 8.

\subsection{Datasets}

% \subsubsection{Base model training dataset}
The \xx{gru} and DPRNN based models are trained using the Deep Noise Suppression (DNS) Challenge dataset~\cite{reddy2020interspeech}. The dataset contains 760.5 hours of clean speech from over 6,000 speakers and a  noise set of over 65,000 clips that cover more than 150 acoustic classes. All speech and noise signals are resampled at 16 kHz. Noisy mixtures are generated on the fly to improve the data diversity, with the \xx{snr} randomly sampled from the range of [-5, 20] dB.

% \subsubsection{Adaptation and evaluation datasets}
The WSJ0 speech corpus~\cite{paul-baker-1992-design} and the WHAM! noise dataset~\cite{wichern2019wham} are used to simulate acoustic scenes for model adaptation and performance evaluation. 
Noise samples are  taken from the WHAM! evaluation subset and  consist of recordings in non-stationary ambient environments such as coffee shops, restaurants, bars, office buildings, parks. To ensure consistent noise scenario in the adaptation and evaluation for one acoustic scene, noise clips from recordings made on the same day and at the same location are selected for simulating the noisy speech in each adaptation and evaluation pair, but with different clips used for each stage to avoid overlap. Following this procedure, we construct 37 distinct noise scenarios. For each noise scenario, mixtures are generated for three separate  \xx{snr} ranges: [-8, 0] dB, [0, 5] dB, and [5, 10] dB, representing very noisy, moderately noisy, and relatively quiet conditions. This results in a total of $37 \times 3 = 111$ simulated testing acoustic scenes. Within each scene, between two and five speakers are randomly selected from the WSJ0 speaker-dependent dataset. Speech samples from the training subset are used for adaptation, while speech samples from the evaluation subset are reserved exclusively for testing.

In the test set, each scene contains 20 samples, so that after mixing, $111 \times 20 = 2220$ clean-noisy pairs are available to evaluate the model performance. For the adaptation stage, the input mixtures are generated following the process proposed in Sec.~\ref{sec:adpt}, for an \xx{snr} range of [-5, 5]\,dB. In addition, both speech and noise signals are randomly cropped into 2\,s segments before mixing. Similar as in  training, all audio files in the adaptation and evaluation are sampled at 16 kHz.

\subsection{Training and Adaptation Details}

Both base models are trained in a supervised manner by minimizing the \xx{mse} between the estimated and target spectrograms. The Adam optimizer is used with an initial learning rate of 1e-3. If the mean training loss does not decrease for two consecutive epochs, the learning rate is reduced by a factor of 10. The batch size in this stage is set to 8, and both models are trained for 100 epochs. During adaptation, we use the inverse \xx{snr} as loss function. The trainable parameters are optimized using Adam with a fixed learning rate of 1e-3 and 5e-4 for \xx{gru}-based models and DPRNN-based models, respectively. The batch size is set to 24, and each adaptation session is limited to 20 updates, corresponding to a maximum utilization of $24\times20=240$ audio signals, which will be 480 seconds.

\begin{table}[t]

\centering
\caption{Performance of adapted GRU models with varying ranks and scaling factors. Results are averaged across all test scenes.}

\vspace{-2mm}
\setlength{\tabcolsep}{5.5pt}
% \resizebox{\columnwidth}{!}{
\begin{tabular}{lcccc}
\toprule
(Rank, Scale) & Adaptable Params & PESQ & STOI & SI-SDR \\
\midrule \midrule
Noisy       & -     & 1.20 & 79.54 & 2.04 \\ 
Pretrained  & -     & 1.36 & 82.28 & 6.75 \\ \midrule
(16, 1)   & 8,192  & 1.42 & 82.59 & 7.85 \\
(32, 1)   & 16,384 & \textbf{1.43} & 82.88 & 7.91 \\
(64, 1)   & 32,768 & \textbf{1.43} & \textbf{82.95} & \textbf{8.03} \\ \midrule
(1, 32)  & 512 & 1.42 & 82.81 & 8.04 \\
(1, 64)  & 512 & \textbf{1.44} & \textbf{82.89} & \textbf{8.17} \\ 
(1, 128) & 512 & 1.41 & 82.65 & 8.14 \\
\bottomrule
\end{tabular} 
\vspace{-3mm}
\label{tab:rank_scale}
\end{table}

\section{Results}
\label{sec:results}

\subsection{ Comparing with SOTA Adaptation Framework }

We compared our proposed lightweight adaptation framework with the state-of-the-art RemixIT method~\cite{tzinis2022remixit}, which requires updating all parameters of the base model. Evaluation was conducted using PESQ~\cite{rix2001perceptual}, STOI~\cite{taal2011algorithm}, and SI-SDR~\cite{le2019sdr} across three SNR ranges. The detailed results are summarized in Table~\ref{tab:adapt_results}.

We first evaluated adaptation in isolated scenes, where each test scene is processed independently and all models were initialized from the DNS-pretrained backbone. As shown in Table~\ref{tab:adapt_results}, both RemixIT and our method improved performance over the pretrained models across all SNR ranges. Importantly, while RemixIT required updating the entire parameter set (230k for GRU, 89k for DPRNN), our method achieved comparable or better gains while updating fewer than 1\% of the parameters (512 for GRU, 708 for DPRNN). For example, with the GRU backbone at [0, 5] dB SNR, our method reached 1.44 PESQ, 85.34 STOI, and 8.36 dB SI-SDR, outperforming RemixIT (1.39 PESQ, 84.96 STOI, 8.27 dB SI-SDR). 
These results demonstrate that parameter-efficient adaptation can deliver performance on par with or exceeding full fine-tuning.
Moreover, to achieve such results, RemixIT requires storing a separate copy of the pretrained model to generate pseudo-targets for self-supervised adaptation, whereas our method only needs to store the lightweight adapter parameters.

Fig.~\ref{fig:snr_improvement} further shows the $\Delta$\xx{snr} in dB of adapted models compared to the pretrained baselines for both \xx{gru} and DPRNN backbones under RemixIT and our framework, across three SNR ranges. With RemixIT, the mean $\Delta$\xx{snr} rises quickly during the initial steps but subsequently oscillates, indicating unstable adaptation dynamics. In contrast, our framework yields a monotonic improvement curve, reflecting stable and consistent gains.

We then created sequential scenes by randomizing all 111 test environments and requiring the model to adapt continuously across them. In this setup, the initialization for each scene comes from the adapted model of the previous scene, simulating long-term deployment. Under this setting, the advantage of our framework becomes more evident. While RemixIT showed degraded performance due to cumulative parameter drift, our method maintained or improved over the pretrained baseline across all metrics. With GRU, our framework achieved 1.72 PESQ, 91.64 STOI, and 11.89 dB SI-SDR at [5, 10] dB, notably higher than RemixIT (1.51 PESQ, 88.63 STOI, 11.03 dB SI-SDR). A similar trend was observed for DPRNN, where our method consistently outperformed RemixIT while adapting only 0.79\% of the parameters.

\subsection{Effect of Rank and Scaling Factor}

Table~\ref{tab:rank_scale} reports the results of models adapted  for different rank and scaling factor configurations when using  the \xx{gru} network as the backbone.  Results are averaged across all test  environments.
Compared to  noisy and pretrained baselines, all \xx{lora}-based adaptations yield consistent improvements in PESQ, STOI, and \xx{sisdr}.
When varying the rank with a fixed scaling factor of 1, performance improves steadily as the rank increases, with the best results achieved at rank 64. However, this comes at the cost of a substantial increase in trainable parameters.
In contrast, varying the scaling factor with a fixed rank of 1 shows that large scaling factors can be far more parameter-efficient: with only 512 trainable parameters, the configuration (1, 64) achieves the best overall results.

\section{Conclusion}

We formalized the problem of post-deployment \xx{se} model adaptation under realistic acoustic environments, which involve dynamic scene changes and demand lightweight solutions. 
To address this challenge, we introduced a self-supervised low-rank adaptation framework that only needs to update fewer than 1\% of the \xx{se} model’s parameters. 
Evaluated across 111 noisy environments, the framework consistently improves PESQ, STOI, and \xx{sisdr}, while maintaining stable progress at each adaptation step. 
These results highlight the practicality of our method for effective and efficient adaptation of speech enhancement models in real-world acoustic conditions.

\vfill\pagebreak

\bibliographystyle{IEEEbib}
\bibliography{refs}

\end{document}